\documentclass[namedreferences]{solarphysics}
\usepackage[optionalrh,solaenum]{spr-sola-addons} 
\usepackage{graphicx}                    
\usepackage[usenames]{color}                       
\usepackage{url}                         

\usepackage{solaheader}	
\newcommand{\solaaccepteddate}{14 Sep 2012}
\newcommand{\Xdoi}{10.1007/s11207-012-0138-y}	%

\newcommand{\arc}{$^{\prime\prime}$}
\newcommand{\vsmr}{VSM/Rockwell}
\newcommand{\vsms}{VSM/Sarnoff}

\newcommand{\aap}{    {\it Astron. Astrophys.}}

\newcommand{\apj}{    {\it Astrophys. J.}}
\newcommand{\apjs}{    {\it Astrophys. J. Supplements}}
\newcommand{\apjl}{    {\it Astrophys. J. Letters}}

\newcommand{\solphys}{{\it Solar Phys.}}

\begin{document}

\begin{article}

\begin{opening}

\title{Comparison of Ground- and Space-based Longitudinal Magnetograms}

%
\author{A. Pietarila $^{1}$\sep      
L. Bertello $^{1}$\sep     
J.~W. Harvey $^{1}$\sep      
A.~A. Pevtsov $^{2}$ }

%

%
  \institute{$^{1}$ National Solar Observatory, 950
  N. Cherry Avenue, Tucson, AZ 85719, USA
                     email: \url{apietarila@nso.edu} 
 $^{2}$ National Solar Observatory, 3010 Coronal Loop
Sunspot, NM 88349, USA            }

\begin{abstract}

We compare photospheric line-of-sight magnetograms from the Synoptic Long-term Investigations of the Sun (SOLIS) vector spectromagnetograph (VSM) instrument with observations from the 150-foot Solar Tower at Mt. Wilson (MWO), 
Helioseismic and Magnetic Imager (HMI) on {\it Solar Dynamics Observatory} (SDO), and Michelson Doppler Imager (MDI) on {\it Solar and Heliospheric Observatory} (SOHO). We find very good agreement between VSM and the other data sources for both disk-averaged flux densities and pixel-by-pixel measurements. We show that the VSM mean flux density time series is of consistently high signal-to-noise with no significant zero-offsets. We discuss in detail some of the factors -spatial resolution, flux dependence and position on the solar disk- affecting the determination of scaling between VSM and SOHO/MDI or SDO/HMI magnetograms. The VSM flux densities agree well with spatially smoothed data from MDI and HMI, although the scaling factors show clear dependence on flux density. The factor to convert VSM to HMI increases with increasing flux density  (from $\approx$1 to $\approx$1.5). The nonlinearity is smaller for the VSM vs. ~SOHO/MDI scaling factor (from $\approx$1 to $\approx$1.2).
\end{abstract}

%
\keywords{Solar magnetic fields, photosphere, magnetograms}

\end{opening}

\section{Introduction}

Synoptic observations play a key role in, {\it e.g.}, space weather
forecasting and solar-stellar studies as well as understanding the
fundamental properties of solar activity. A good synoptic data set has
a long duration and consistently high quality {\it e.g.}, no artificial trends or jumps in the data and only few outlier points. For long-term
measurements, however, it is inevitable that instruments will be
upgraded or even replaced. It is therefore critical to ensure the
continuity of data sets from one instrument to
another. Inter-instrument comparisons are one way to ensure the
continuity. The aim of this paper is to provide a comparison between the
Synoptic Long-term Investigations of the Sun (SOLIS;
\opencite{Keller+others2003}) photospheric line-of-sight (LOS) magnetograms and the
synoptic data sets from the 150-foot Solar Tower at Mt. Wilson (MWO; \opencite{Ulrich+others2002}), Helioseismic and Magnetic Imager (HMI; \opencite{HMI}) on the {\it Solar Dynamics Observatory} (SDO) satellite and Michelson Doppler Imager (MDI; \opencite{mdi}) on the {\it Solar and Heliospheric Observatory} (SOHO) satellite.

In earlier days of solar magnetography, several studies were conducted
to compare the LOS and vector magnetograms taken by different
observatories. These early studies indicated difficulties when
comparing the not strictly simultaneous observations taken with
different spatial and spectral resolution and in different spectral
lines. The results of comparisons also depend on position on the solar disk and which solar feature is under consideration (see \inlinecite{Demidov+others2008} for details and additional references). More recently, \inlinecite{Leka+Barnes2012} present a
comprehensive study of the effects of different spatial resolutions on
magnetic field measurement comparisons confirming many earlier
findings. Other examples of magnetogram comparisons include
\citeauthor{Tran+others2005} (\citeyear{Tran+others2005}; SOHO/MDI and MWO), \citeauthor{Berger+Lites2003} (\citeyear{Berger+Lites2003}; Advanced Stokes Polarimeter at NSO/Sacramento Peak and  and SOHO/MDI), and \citeauthor{Liu+others2012} (\citeyear{Liu+others2012}; SDO/HMI and SOHO/MDI). While the three comparisons above include data from
either MDI, MWO, and/or HMI, the main difference between
them and the comparison presented here is that they are either for
very low spatial resolution, all space-based instruments, or for a
single observation. In contrast, SOLIS magnetograms are synoptic,
ground-based, and have moderate spatial resolution. The vector spectro-magnetograph (VSM) on SOLIS was designed in part to continue the synoptic data set of LOS magnetograms from the 512-channel magnetograph and the NASA/NSO Spectromagnetograph (SPM) operated during 1974--2003 at NSO/Kitt Peak (\opencite{kpvt1},\citeyear{kpvt2}; \opencite{Jones+others1992}). Comparison of SOLIS/VSM measurements with MDI, HMI, and MWO is necessary for ensuring the continuity of the combined data set from these two NSO instruments.

\section{SOLIS VSM}

SOLIS is a synoptic facility to study the solar activity cycle,
energetics in the solar atmosphere, and solar irradiance changes. It
consists of three telescopes: The 50 cm vector
spectro-magnetograph (VSM), the 8 mm integrated sunlight
spectrometer (ISS), and the 14 cm full-disk patrol
(FDP). Currently there is one SOLIS system operating at Kitt Peak in
Arizona, but a global network of SOLIS/VSM instruments is envisioned in the
NSO long range plan. VSM, operational since 2003, is designed to
measure solar magnetic fields in a synoptic manner to address the
SOLIS science questions and to continue the synoptic observations made
by NSO during 1974-2003. The VSM magnetic field
observations include measurements of the full vector field (Stokes
$I$, $Q$, $U$ and $V$) using the photospheric Fe {\sc i} 630.15 and
630.25 nm lines and full-disk LOS flux density measurements (Stokes $I$
and $V$) using photospheric Fe {\sc i} 630.15 nm and chromospheric
Ca {\sc ii} 854.2 nm lines.

 Several authors ({\it
  e.g.}, \opencite{Jones+Ceja2001}; \opencite{Thornton+Jones2002};
\opencite{Wenzler+others2004}; \opencite{Demidov+others2008}) have
shown that the Kitt Peak SPM and MDI data are in good agreement except
that the MDI measurements are a factor $\approx$1.4 higher than raw SPM
measurements
\cite{Jones+Ceja2001,Thornton+Jones2002,Wenzler+others2004}. \inlinecite{Jones+others2004}
compared VSM measurements with SPM and found that the two are
virtually identical except for the improved sensitivity of VSM. Since
the agreement between SPM and MDI as well as SPM and VSM has already
been established by other authors, we do not include SPM data in the
current analysis. The differences between measurements can be
attributed to several factors. \citeauthor{Berger+Lites2002} (\citeyear{Berger+Lites2002}, \citeyear{Berger+Lites2003}) present a
detailed comparison of magnetograms derived from slit- and
filter-based instruments and conclude that the response of filter-based
magnetograms depends strongly on thermodynamic conditions,
spatial resolution and field inclination effects. The effect of
spectral resolution on LOS flux measurements and choice of method used
to measure the flux was addressed by \inlinecite{Cauzzi+others1993}. They
found that the center of gravity method \cite{Rees+Semel1979} offers
the best way to measure LOS fields from high and moderate spectral
resolution observations while the derivative method based on the weak field approximation and the single wavelength modeling-based calibration result in increased scatter. Also other factors such as
calibration procedures, {\it e.g.,} treatment of instrumental
polarization and zero-offsets, and choice of spectral line can
contribute to discrepancies between instruments. Due to these multiple
factors even spectrograph-based instruments do not necessarily result
in identical magnetograms.

A detailed description of the VSM telescope and spectrograph can be
found in, {\it e.g.}, \inlinecite{solis} and
\inlinecite{Keller+others2003}. The VSM magnetograms are constructed
by scanning (2048 steps) in declination (from geocentric south to
north) with the spectrograph slit across the solar disk. The slit,
2048\arc\ in length, is oriented in the east-west direction. Thus, a
single scanline provides full spectra for all points on the solar disk from the eastern to the western limb. Currently there is no telescope
guider to stabilize the image during the observations. This may lead
to shifts of the solar image relative to the slit during bad seeing
conditions. Seeing conditions can be considered bad when granulation
is not visible in the continuum intensity images. To account for the
curvature of the spectrum, the slit is curved with a radius of
16173\arc\ (the center of the slit is shifted southward by
approximately 32\arc\ relative to the endpoints of the slit). The
image of the spectrum is split by a focal plane beam-splitter into two
parts, each 1024\arc\ long, which are re-imaged onto two separate
1024$\times$256 pixel CCD cameras (camera A and camera B). The LOS
magnetograms are derived from the Stokes ($I+V$) and ($I$-$V$) spectra
(nominal spectral sampling is 23 m\AA) using the SPM algorithm
\cite{Jones+others1992,Jones1996}, which is a variant of the center of
gravity method \cite{Rees+Semel1979}.

The computed magnetograms have an instrumental zero-point magnetic
field (``magbias'')  that varies as a function of pixel position along the slit. The variations
are typically within $\pm$ 2 G (gauss) for the Fe {\sc i} 630.15 nm line. To remove the magbias, we
 compute the flux distribution of pixels located in vertical bands (parallel to  the scanning direction)
across the solar image. Each band is 11 pixels wide and displaced by 1 pixel with respect to the preceding
band. Only pixels with values in the range $\pm$ 15 G are used.
To mitigate the effect of decaying active regions each band includes pixels from the current observation
and also from observations taken during  the previous 14 days. A Gaussian fit is applied
to determine the center (magbias value) of the individual distributions. The magbias is subtracted from the magnetograms. The end product, zero-point corrected magnetograms, are geometrically corrected to remove the slit curvature and to merge the
cameras (by removing the gap between the two halves of the solar disk as observed by the two cameras) in order to produce a circular solar disk with a spatial
sampling of about 1\arc.

The major changes to VSM since 2003 include a move of the telescope
from a temporary site to its current location on Kitt Peak in
2004. The original VSM modulators were replaced in 2006 leading to
better polarimetric measurements.  The original Rockwell cameras had a
pixel size of 1.125\arc. In 2010 they were replaced by Sarnoff cameras
with smaller pixel size, 1\arc, and better sensitivity. The circular
polarization efficiency of the current LOS modulator was carefully
measured in 2011 and found to range from 0.96 at the east end of the
slit to 0.75 at the west end. The modulator efficiency was
measured by inserting a linear polarizer followed by a quarter-wave
plate in front of the telescope focal plane. Sixty-four sets of solar
spectra were recorded with various angles of the polarizer and wave
plate relative to the slit direction. Each spectral set consisted of
two simultaneous spectra produced by a Savart plate polarizing beam
splitter located just in front of the camera focal plane and two
sequential spectra produced by changing the modulator state. At each
slit position the observed intensities of the four spectra for all 64
combinations of polarizer and wave plate angles were fit to a model of
the polarization properties of the telescope and spectrograph. Seven
parameters of the model were allowed to vary in the fitting process
while another seven were fixed based on previous measurements. One of
the adjustable parameters is the amount by which circular polarization
fails to be completely modulated as recorded by the cameras. This
factor varies along the slit and depends mainly on imperfect
performance of the modulator, unmodulated scattered light within the
spectrograph, and image lag within the camera CCD detectors. Dividing
one by the varying factor produces a multiplicative correction
function along the slit that is applied to recorded magnetograms. This
process corrects measured magnetic field values that are weakened by
internal instrumental effects. Calibration data needed to construct a
correction function are not available for the Rockwell era. 

The level
of scattered light in VSM is less than 10\%. It varies with position
in the final optical plane and with the position of the Sun on the
entrance slit. The correction function for the efficiency of the modulator for Sarnoff 
includes instrumental scattered light. Scattered light from other sources, {\it
i.e.}, atmosphere, telescope and image blurring, is not accounted
for. The reduction in the apparent circular polarization signal due to
scattered light is most pronounced in sunspot umbrae.

An internal calibration to account for the camera change was done by comparing the widths of flux density distributions in remapped heliographic (180$^{\circ} \times$180$^{\circ}$) magnetograms during 2009 and 2010. It was determined that a scaling factor of 1.59 needs to be applied to the raw \vsmr\ magnetograms to reconcile the observations made with \vsmr\ and \vsms. This relatively large factor is attributed to image lag in the Rockwell cameras and not using a modulator efficiency correction function for Rockwell data.

The full-disk LOS Fe {\sc i} 630.15 nm magnetograms (referred to as Level2 data
products) are used to create synoptic maps and mean magnetic flux density time
series (Level3). All SOLIS data products are available on-line at: \url{http://solis.nso.edu}. For the comparisons discussed in this paper we use VSM data taken after the new modulator was installed in 2006.

\subsection{Weak Signal Distribution}

\begin{figure*}
\includegraphics[width=12cm]{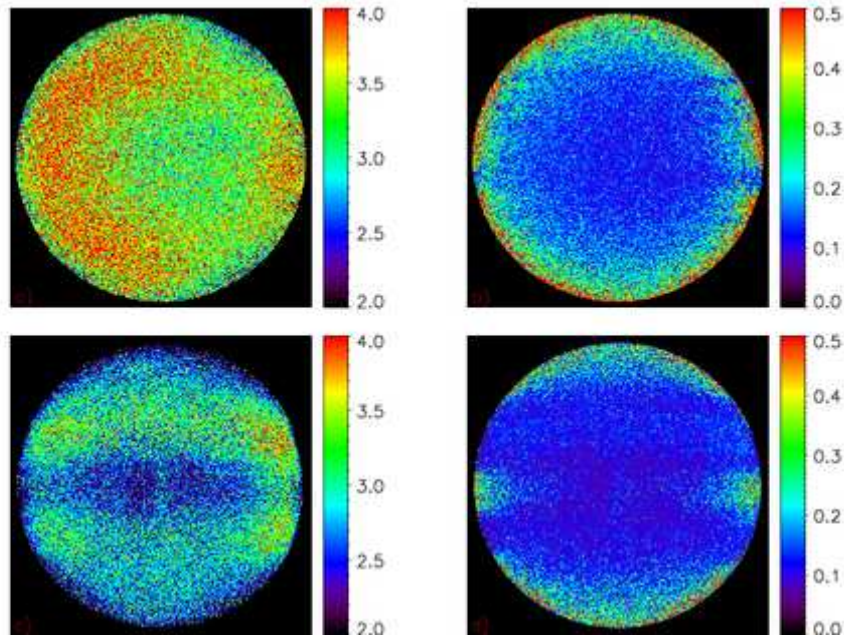}
\caption{Spatial variation of noise in VSM full
disk LOS magnetograms. (a) \vsmr\ noise in units of G measured as the standard deviation, $\sigma$, of
a fitted Gaussian distribution. (b) one-$\sigma$ error estimates for
the Gaussian fits. Panels (c) and (d) are the same as (a) and (b) but for \vsms.}
\label{fig:noise}
\end{figure*}

To estimate the noise level of VSM LOS Fe {\sc i} 630.15 nm magnetograms and how it varies spatially
we make a data cube ($x, y, n$) of all magnetograms where the solar radius is
larger than 855 pixels (for \vsmr, 272 magnetograms) or 970 pixels (\vsms, 195 magnetograms). The radius criterion is used to
minimize the effect of size variation of the solar disk on the noise levels per pixel. We then fit Gaussian functions to
the distributions of low flux density pixels (below 100 G) for each pixel on the
solar disk. The width, $\sigma$, of the Gaussian is used as a proxy for noise. This is an
upper limit for instrumental noise since the distribution is dominated by a significant amount of genuine weak solar signals: Measurements of the instrumental noise using out of focus images give a pixel-by-pixel noise level of well below 1 G. The
results of the analysis are shown in figure \ref{fig:noise}.

The \vsms\ noise image (Figure 1c) also includes visible
contributions from solar activity, not only weak quiet Sun fields, as seen by the presence of activity belts in the
noise images. The mean noise is 2.8 G in \vsms\ and slightly higher, 3.5 G,
in \vsmr. In both cases the noise is lowest at disk center and increases
toward the limb. The noise appears to decrease again very near the limb. The errors
of the Gaussian fits, however, also increase near the limb making it difficult to judge the reason for the decrease. In fact, the near limb behavior is expected for a distribution of nearly horizontal fields that become difficult to resolve near the limb (see \opencite{Harvey+others2007}). Both \vsmr\
and \vsms\ noise images have an area of increased noise near the west
limb. This is likely caused by minor vignetting and consequent light level reduction near the west edge of the slit. There is
an additional east-west hemisphere asymmetry in the \vsmr\ magnetograms, with
the exception of the west limb, camera B has lower noise than camera A. This may be due to variation in modulator efficiency (modulator inefficiency is corrected for only in \vsms\ magnetograms) and possible differences in gain and scattered light. \vsms\ data do not show this asymmetry. The area around the gap
between camera A and camera B is visible as a region
of increased noise. The prism that splits the spectrum into the camera A and camera B beams has some dust particles on it. These are nearly in focus at the vertex of the prism but rapidly defocus away from the vertex. When the dust is in focus, it adds noise to the algorithm used to determine the LOS flux density. Since the gap is not always placed on the same position
relative to the solar disk ($P$-angle variations and centering of the solar image on the slit) the width of the increased noise region may be exaggerated in the noise
image.

\section{Comparison of VSM Magnetograms with Other Instruments}

\subsection{Mean Magnetic Flux Density Comparison}

\begin{figure*}
\includegraphics[width=12cm]{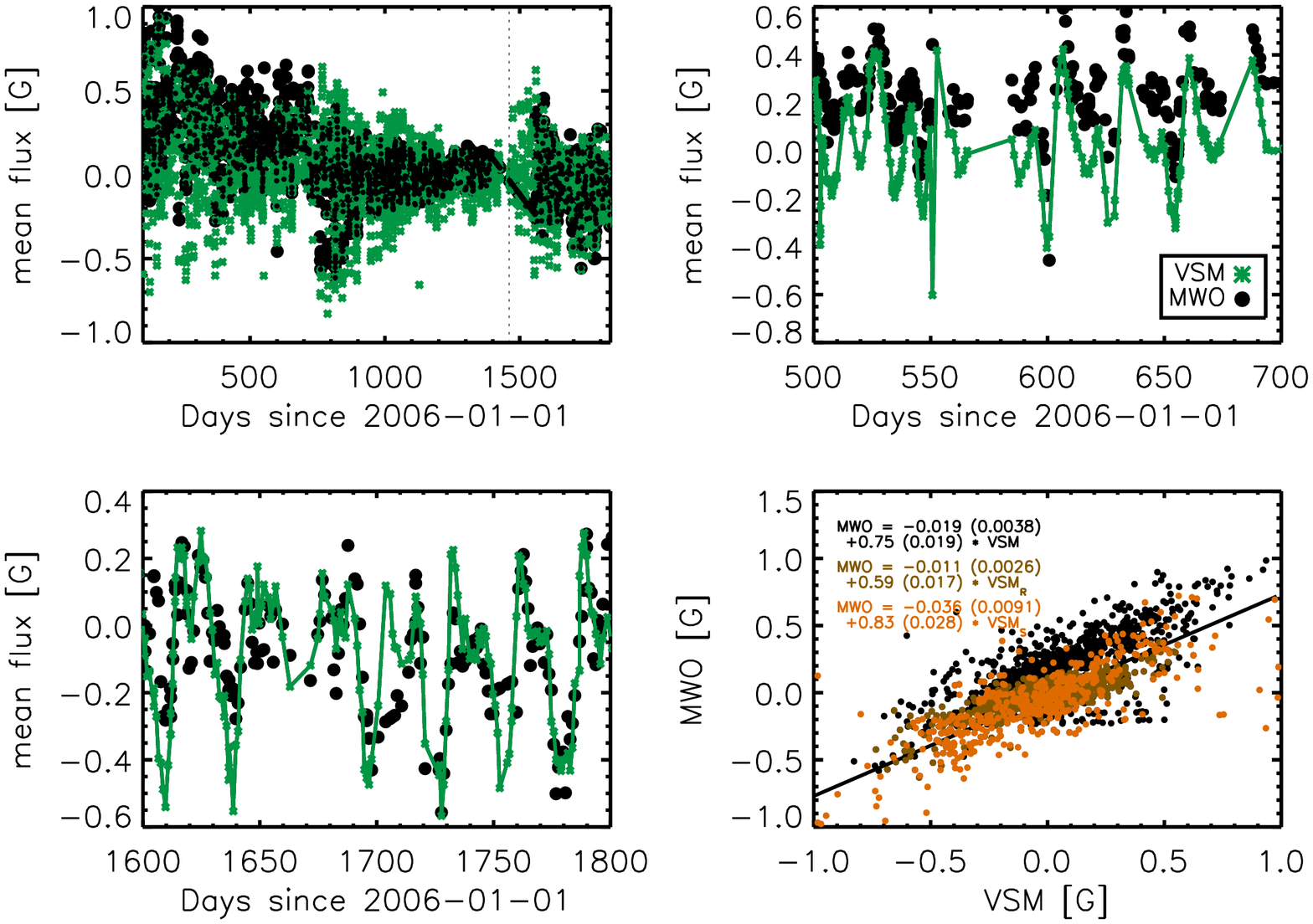}
\caption{Comparison of VSM and MWO mean magnetic flux density
measurements. Top left: VSM (green) and MWO (black) mean magnetic flux density as a
function of time. Vertical dotted line marks change from \vsmr\ to \vsms. Top right and bottom left: Magnifications of portions of the
time series. Bottom right: Scatter plot of VSM vs. MWO mean flux density
measurements. All data points are in black,
\vsmr\ (excluding the first 780 days) in brown, and \vsms in orange. Text indicates results of linear fits to the scatter plots with
the same color-coding as used for the data points. Standard deviation
of the fit parameters is given in parentheses. }
\label{fig:wilson}
\end{figure*}

\begin{figure*}
\includegraphics[width=12cm]{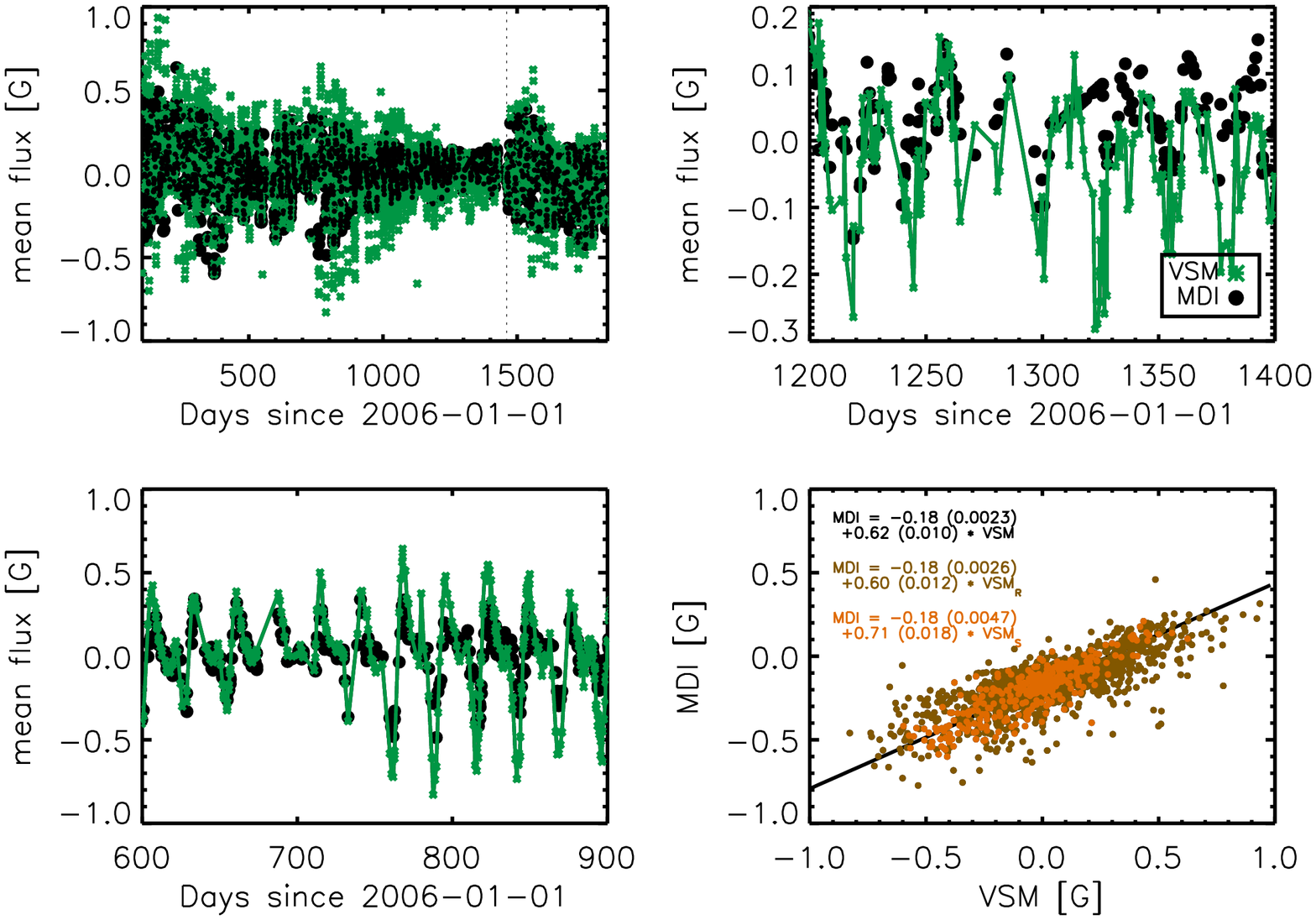}
\caption{Comparison of VSM and MDI mean magnetic flux density
measurements. Top left: VSM (green) and MDI (black) mean magnetic flux densities as a function
of time. Vertical dotted line marks change from \vsmr\ to \vsms. Top right and bottom left: Magnifications of portions of the
time series. Bottom right: Scatter plot of VSM vs. MDI mean flux density
measurements. Data points from \vsmr\ are in brown and \vsms\
data points are in orange. Text indicates results of linear fits to the scatter plots
with the same color-coding as used for the data points. Standard
deviation of the fit parameters is given in parentheses.}
\label{fig:mdi}
\end{figure*}

We compare VSM mean flux density measurements (flux density averaged over the entire disk) with MDI and MWO (R. K. Ulrich, private communication)
data. The
MDI and MWO data are interpolated to the same temporal sampling as VSM and
all pixels with values above 1 G are excluded from the analysis. Figures \ref{fig:wilson} and \ref{fig:mdi}
and Table \ref{tab:mf} summarize the results of the comparison. The first 780 days of the MWO mean flux density time series have a
different zero-offset than the rest of the time series (see the top panels of figure
\ref{fig:wilson}). We exclude these data points from the analysis.

\begin{table}
\caption{Summary of mean flux density comparisons. The first column denotes which data sets are compared. $a$ and $b$ are coefficients of a linear fitting: MDI (MWO) = $a + b\times$ VSM. The standard deviation of the fit parameters is given in parentheses. $r_{\rm c}$ is the Pearson correlation coefficient. }
\label{tab:mf}
\begin{tabular}{llll}
\hline
 & $a$  & $b$  &$r_{\rm c}$ \\
MWO vs. ~VSM & -0.019 (0.0038) & 0.75 (0.019) & 0.77 \\
MWO vs. ~\vsmr &  0.0011 (0.0026) & 0.59 (0.017) & 0.80 \\
MWO vs. ~\vsms & -0.0035 (0.0091) & 0.83 (0.028) & 0.77 \\
MDI vs. ~VSM & -0.18 (0.0023) & 0.62 (0.010) & 0.78 \\
MDI vs. ~\vsmr & -0.18 (0.0020) & 0.60 (0.012) & 0.76 \\
MDI vs. ~\vsms & -0.18 (0.0040) & 0.71 (0.018) & 0.86 \\
\end{tabular}
\end{table}

The overall agreement between
the data sets, VSM, MWO and MDI, is good: The change in activity level and
solar rotation modulation in the three time series are similar and the correlation
between the different series is strong. A linear bisector fit \cite{sixlin}, MDI (MWO) =
$a+b \times$ VSM, gives scaling factors of approximately 0.6
for VSM vs. ~MDI and 0.75 for VSM vs. ~MWO.

The zero-offsets for the MWO and VSM mean fluxes are small. For MDI,
a considerably larger value is found. It is known that MDI level 1.8 data has
a zero-offset so we attribute the offset to be mostly due to MDI. Based on the
analysis, the “magbias” applied to the VSM data successfully removes instrumental
artifacts. No difference in zero-offsets is seen for \vsmr\ and
\vsms.

If the VSM time series is divided into \vsmr\ and \vsms\ and
the comparison with MDI and MWO is repeated, the retrieved scaling factors
for \vsmr\ and \vsms\ are not consistent: The \vsms\
gives higher values. If only weak fields (data points where the mean flux density
is less than 0.3 G) are included, the resulting scaling factors for MDI vs. ~\vsmr\ and MDI vs. ~\vsms\ agree within error bars of the fits, 0.62. The disagreement
of the mean flux density scaling and agreement of the scaling if only weak fluxes are included indicates that the internal scaling of \vsmr\
and \vsms\ is nonlinear, {\it i.e}., weak and strong fields cannot
be scaled with a single factor.

\subsection{Pixel-by-pixel LOS Magnetogram Comparison}

We also make a comparison of VSM magnetograms with HMI and MDI magnetograms
on a pixel-by-pixel basis. All the scaling factors are determined from a linear bisector fit
to the MDI (HMI) and VSM data points: MDI (HMI) = $a + b\times$\vsmr\ (\vsms). The 1-$\sigma$ errors of the measured scaling factors
 are small (less than $10^{-3}$) and not shown for most of the results. 

\begin{figure*}
\includegraphics[width=12cm]{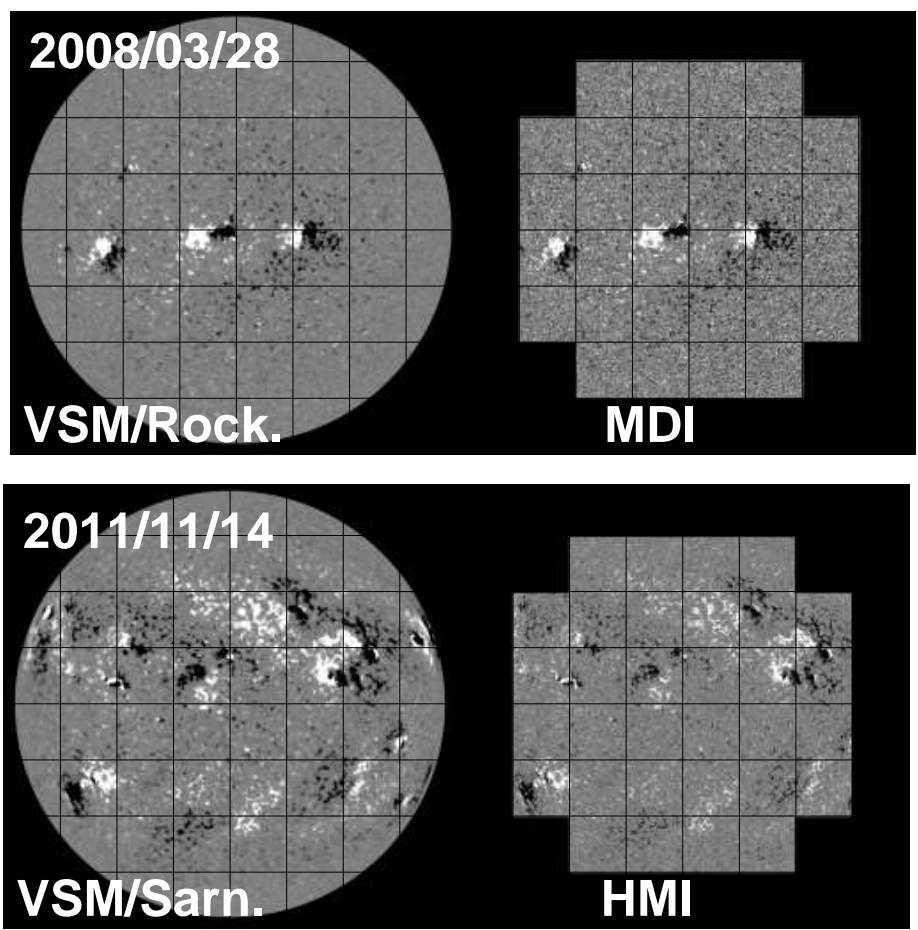}
\caption{Examples of full-disk LOS magnetograms used for pixel-by-pixel
comparisons. The grid shows tiling used for co-alignment. All
magnetograms are scaled to [-50,50] G.}
\label{fig:fd}
\end{figure*}

For the pixel-by-pixel comparison, we use full-disk magnetograms from
MDI (level 1.8) and HMI (45 s). We compare the magnetograms on
the largest pixel scale common to the MDI vs. ~\vsmr\ and HMI vs. ~\vsms\ pairs
(1.98\arc\ for MDI vs. ~\vsmr\ and 1\arc\ for HMI-\vsms). To co-align the
magnetograms we divide all the images in 8$\times$8 tiles
(254$\times$254 and 128$\times$128 pixels for HMI and MDI,
respectively). Only tiles which are fully on the solar disk are
included in the analysis (see figure \ref{fig:fd}). To co-register the
images we use the Solar Software \cite{SSW} image co-alignment routine
auto\_align\_images.pro, which determines the scaling and shifts in
$x$ and $y$ as well as allows for possible warping of the
magnetograms, due to varying seeing conditions within the VSM
data. The alignment is optimized using the Powell minimization
algorithm. Cubic convolution is used to interpolate the images onto a
common grid. The resulting co-alignment is not very sensitive to the
method used: Rigid displacements computed via cross correlations yield
very similar results, although the quality of the co-alignment using
the first method is better near the limb. We co-align two sets of
MDI vs. ~\vsmr\ and three sets of HMI vs. ~\vsms\ magnetograms.

The instruments differ not only by their pixel sizes, but in also
spatial resolutions and point spread functions leading to very different
apparent flux density per pixel of the magnetograms. VSM observations
are affected by variations in seeing conditions which can lead to a
varying spatial resolution in different magnetograms and, at times,
even within a single magnetogram. In particular, the VSM magnetograms
are affected during bad seeing conditions by perturbations in seeing
moving the solar image on the slit. To compensate for the different
spatial resolutions we spatially smooth the space-based images by
convolving them with a Gaussian of a given FWHM (FWHM=2.35$\times
\sigma$). A better match between the magnetograms might have been
achieved through using a two parameter Voigt function. However, to
minimize the number of free parameters in the comparisons, we chose to
use a Gaussian with a varying FWHM. The spatially smoothed
magnetograms have apparent flux densities in better agreement with the
VSM data, but the smoothing also artificially reduces the apparent
flux density of the space-based measurements.

\begin{figure*}
\includegraphics[width=12cm]{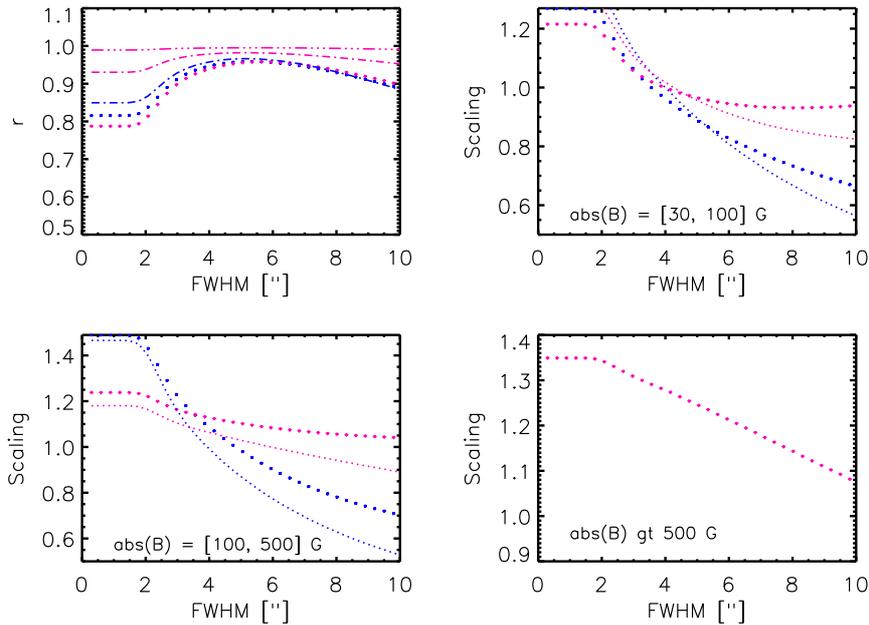}
\caption{Effect of spatial smoothing in determining scaling factors
for MDI vs. ~\vsmr. Upper left: Pearson correlation coefficient of
MDI vs. ~\vsmr\ magnetograms as a function of FWHM of the Gaussian used to
spatially smooth the MDI data. Dotted lines are from pixels with VSM flux density
between 30 and 100 G, dash-dotted lines for 100--500 G and dash-triple
dotted for greater than 500 G. Different colors denote different data
sets. Upper right: Scaling factor as a function of smoothing for
pixels with VSM flux density between 30 and 100 G. Larger dots are for
center tiles and smaller dots for edge tiles. Lower left: Scaling
factor for pixels with VSM flux density between 100 and 500 G. Lower
right: Scaling factor for pixels with flux density greater than 500
G.}
\label{fig:rock-var}
\end{figure*}

\begin{figure*}
\includegraphics[width=12cm]{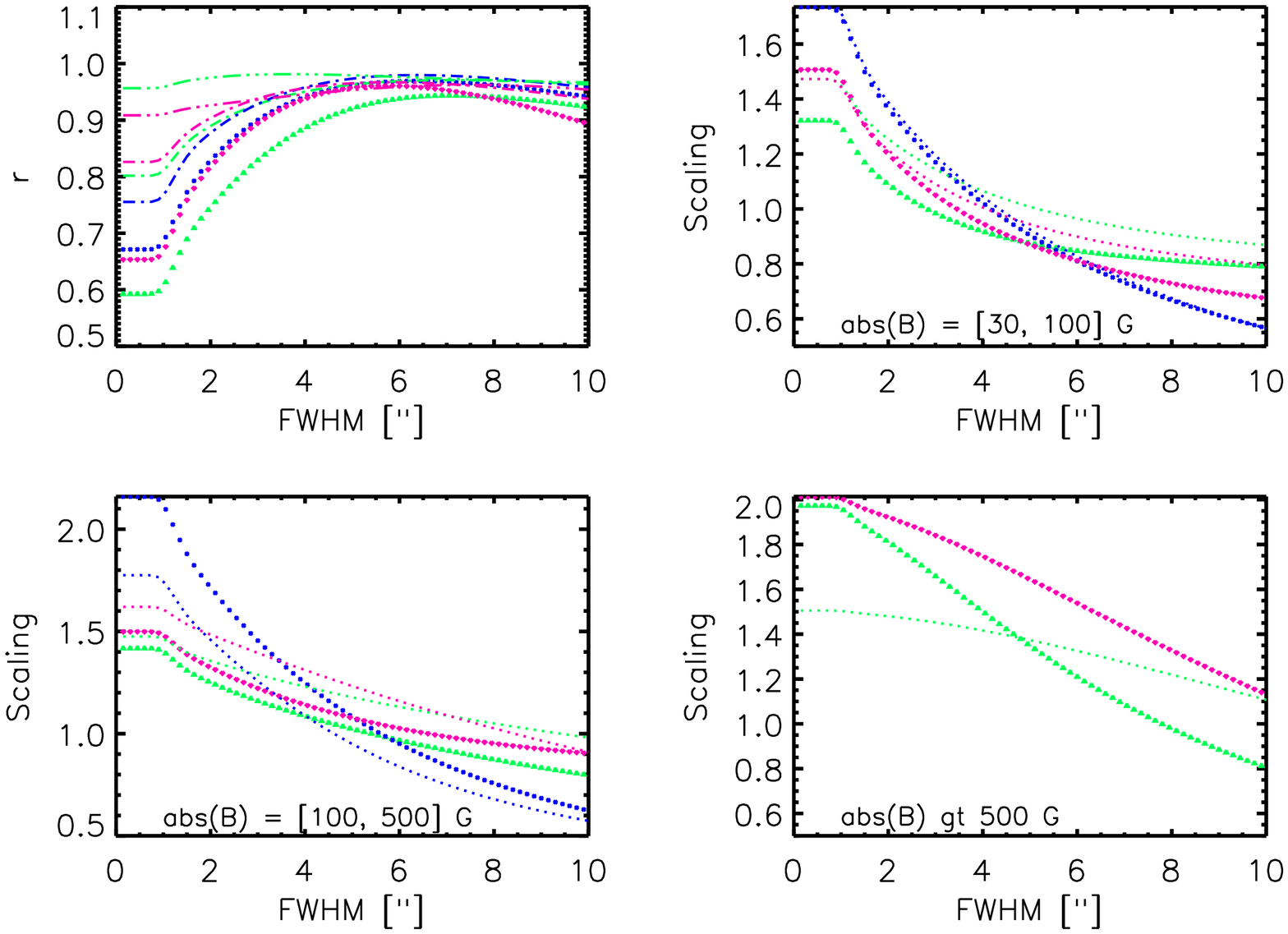}
\caption{Effect of spatial smoothing in determining scaling factors
for HMI vs. ~\vsms. Upper left: Pearson correlation coefficient of
HMI vs. ~\vsms\ magnetograms as a function of FWHM of the Gaussian used to
spatially smooth the HMI data. Dotted lines are from pixels with VSM flux density
between 30 and 100 G, dash-dotted lines for 100--500 G and dash-triple
dotted for greater than 500 G. Different colors denote different data
sets. Upper right: Scaling factor as a function of smoothing for
pixels with VSM flux density between 30 and 100 G. Larger dots are for
center tiles and  smaller dots for edge tiles. Lower left: Scaling
factor for pixels with VSM flux density between 100 and 500 G. Lower
right: Scaling factor for pixels with flux density greater than 500
G.}
\label{fig:sarn-var}
\end{figure*}

Figures \ref{fig:rock-var} and \ref{fig:sarn-var} demonstrate how
spatial smoothing changes the magnetograms. Maximizing the correlation
coefficient between the magnetograms effectively optimizes the
agreement of the apparent flux densities.  However, smoothing affects
the correlation coefficients differently for weak or strong fluxes.

Scaling factors for the unsmoothed MDI data and VSM data are
$\sim$1.2-1.4, similar to the values found for SPM and MDI
\cite{Jones+Ceja2001,Wenzler+others2004}. This further demonstrates
the good agreement between SPM and VSM data as previously shown by
\inlinecite{Jones+others2004} for early VSM data.

For strong fluxes the effect of smoothing on the correlation function
is fairly small. Strong fluxes are easy to co-align and they are
nearly always surrounded by fairly strong fluxes leading to strong
correlation, even in the case of spatially unsmoothed data.  The
scaling factor for the strong fluxes decreases with increased
smoothing.

For intermediate fluxes (typical features are network fields), the
correlation first increases with increased smoothing, reaches a
plateau, and eventually begins to decrease again. The scaling factors
decrease with increased smoothing, but not in as linear a manner as was
the case for the strong fluxes. For the intermediate fluxes an optimal
minimum amount of smoothing ({\it i.e.}, when the plateau is reached)
can be roughly determined from figures \ref{fig:rock-var} and
\ref{fig:sarn-var}: For HMI vs. ~\vsms\ it is 4\arc\ and for MDI vs. ~\vsmr\
5\arc. Note that the optimal smoothing is not the same for all the
data sets. In figure \ref{fig:sarn-var} the green lines correspond to
a set when the VSM magnetogram was taken under seeing conditions worse
than the two other data sets included in the analysis. The initial
correlation coefficient for it is smaller and as a function of
smoothing the curve peaks at higher values than for the two other data
sets.  The effect of location on the solar disk for determining
scaling factors is already implied in figures \ref{fig:rock-var} and
\ref{fig:sarn-var}: Disk center (tiles surrounded in all directions by
tiles fully located on the solar disk) and edge (tiles fully on the
solar disk, but partially surrounded by tiles not fully on the disk)
give slightly different scaling factors.

\begin{figure*}
\includegraphics[width=12cm]{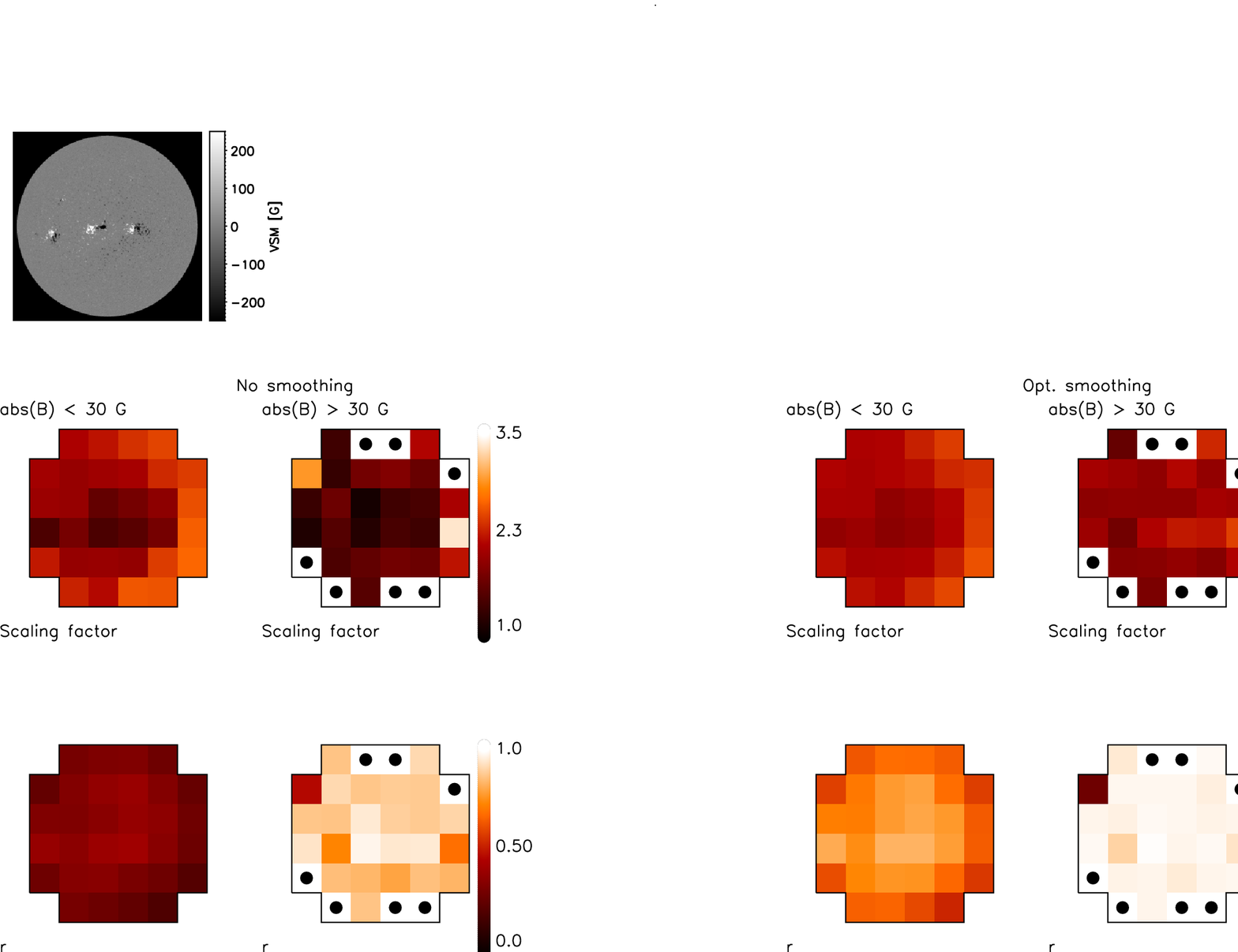}
\includegraphics[width=12cm]{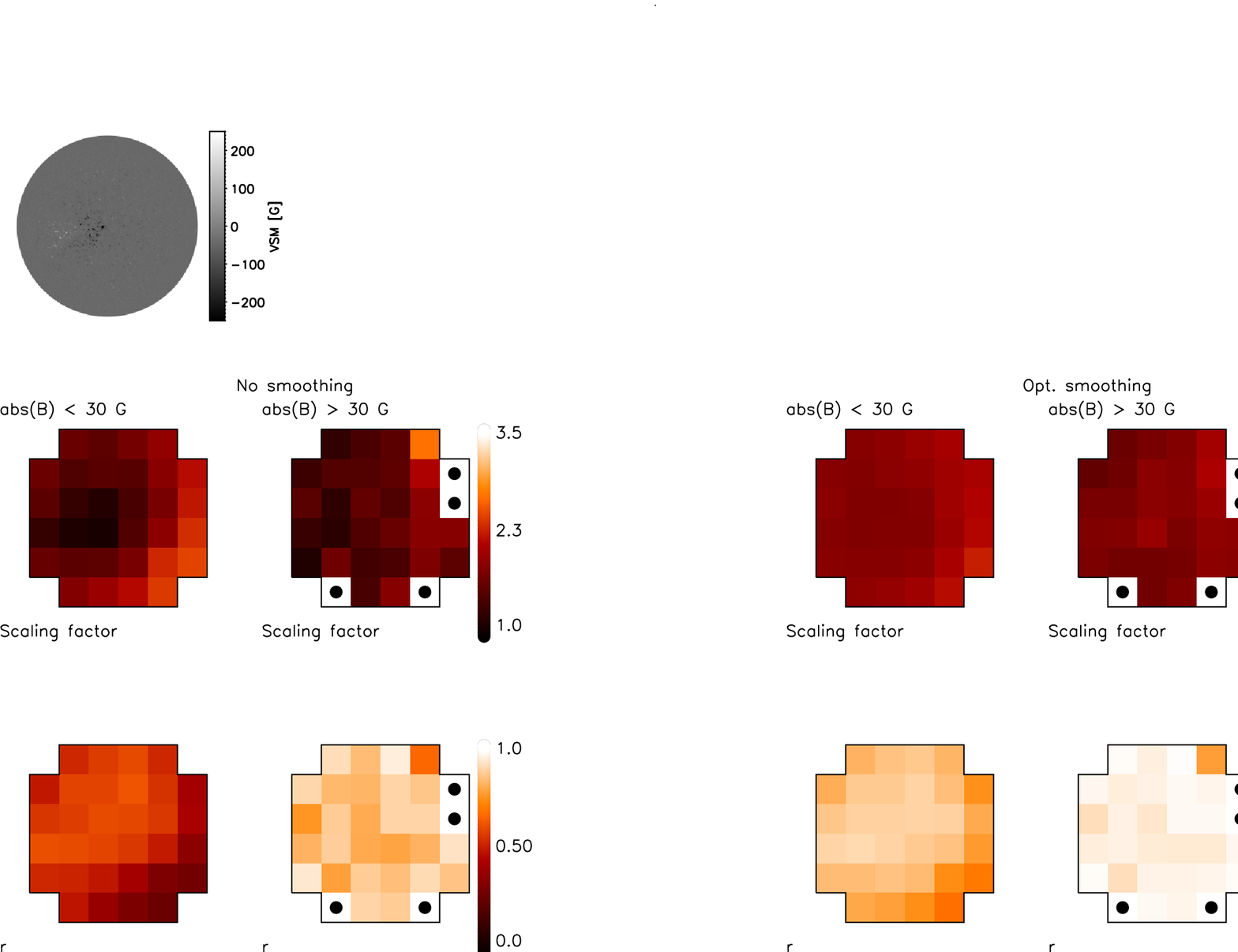}
\caption{Scaling factors and correlation coefficients for two sets of \vsmr\ and MDI magnetograms. Dots indicate tiles for which the statistics is insufficient to determine the scaling factors.  }
\label{fig:mdi-slope}
\end{figure*}

Figure \ref{fig:mdi-slope} shows how the MDI vs. ~\vsmr\ scaling factors
change across the solar disk. Both sets of magnetograms have an
east-west asymmetry in the behavior of weak (below 30 G, likely
dominated by noise) fields: The west side (especially lower right
corner) has weaker correlation and larger scaling factors.  This is
consistent with higher noise in MDI magnetograms in the bottom right
corner; see Figure 9 in \inlinecite{Liu+others2012}. Spatial smoothing
(using the ``optimal'' 5\arc\ FWHM Gaussian) increases the spatial
homogeneity of the scaling factors, especially for fields clearly
above the noise level of unsmoothed MDI data. The spatial variation of
the scaling factors can be attributed to center-to-limb variation (and
spatial asymmetries in instruments) as well as the amount of flux
present in the tile.

\begin{figure*}
\includegraphics[width=12cm]{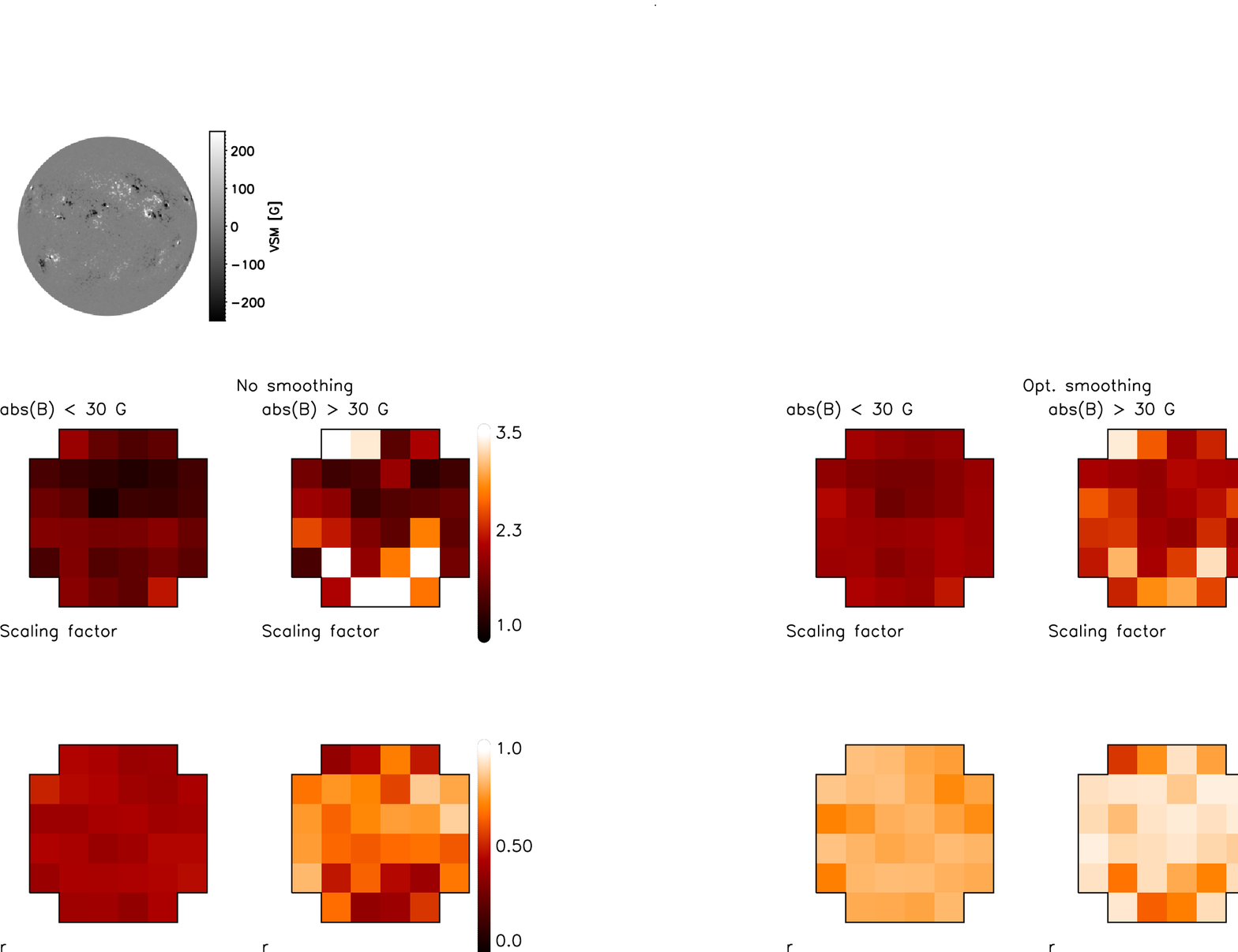}
\includegraphics[width=12cm]{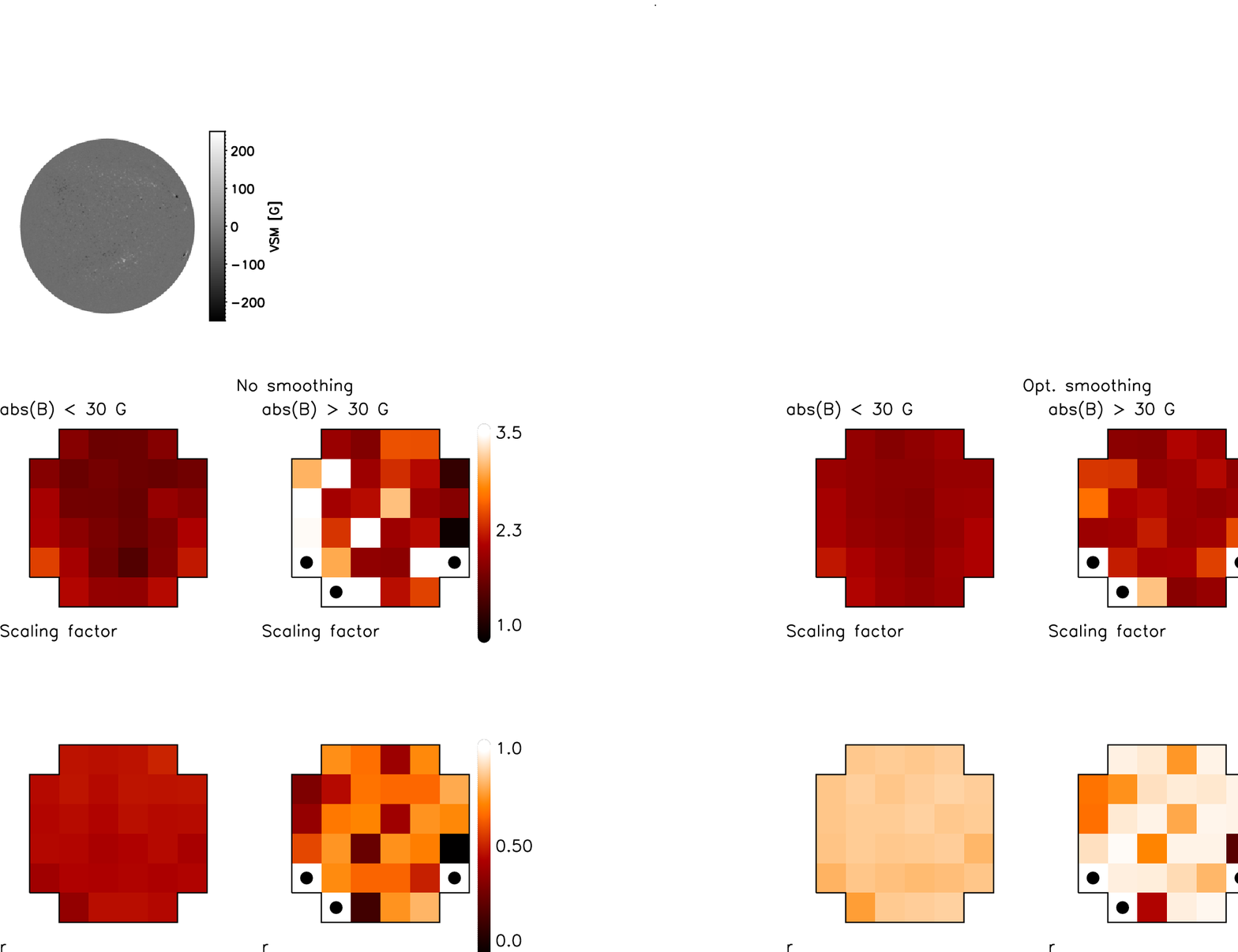}
\caption{Scaling factors and correlation coefficients for two sets of \vsms\ and HMI magnetograms. Dots indicate tiles for which the statistics is insufficient to determine the scaling factors. }
\label{fig:hmi-slope}
\end{figure*}

The same tile-by-tile analysis is shown for HMI vs. ~\vsms\ in figure \ref{fig:hmi-slope}.
Unlike the MDI vs. ~\vsmr\ case, there is no apparent east-west asymmetry. Tiles with low correlation tend to have higher scaling factors. In the smoothed
data these tiles are located mostly near the limb where the noise is higher. The spatial variation of the scaling factors is also partly due to the nonlinearity of the scaling factor as a function of
flux density: The scaling between HMI vs. ~\vsms\ is more nonlinear than for
MDI vs. ~\vsmr\ as shown below.

Figures \ref{fig:mdi-dflux} and \ref{fig:hmi-dflux} demonstrate how the scaling factors depend on the magnetic flux density range
considered. All tiles fully on the solar disk are included in the analysis, {\it i.e.}, no consideration is given to the relative position, edge or center, of the tiles. The
analysis is made for ”optimally” smoothed data (4\arc\ for HMI vs. ~\vsms\ and 5\arc\
for MDI vs. ~\vsmr). Spatial smoothing reduces the noise of the MDI (HMI)
magnetograms and brings them closer to the VSM data. Fitting a Gaussian to the
histograms of the \vsmr\ and \vsms\ magnetograms gives widths, $\sigma$
 of 3.8 G and 3.0 G, respectively. Prior to smoothing MDI and HMI $\sigma$s
are 24.0 G and 8.8 G, and after smoothing 7.5 G and 3.4 G, respectively. Due to the smaller
dynamic range (BSCALE=2.8 G) of MDI, the MDI histogram is binned
in increments of 5 G instead of 1 G as was done for the other magnetograms.
Doing this gives a slightly larger $\sigma$. If the \vsmr\ histogram is binned
in increments of 5 G the resulting Gaussian has a slightly larger $\sigma$, 4.3 G. 

The flux dependence of the scaling factors is systematic: The
scaling factors are larger for higher fluxes. The difference is more modest for
MDI vs. ~\vsmr\ (the scaling factor is $\approx$1.0 for weak fields, $\approx$1.2 for strong fields) than for HMI vs. ~\vsms\ (varies from $\approx$1.0 to $\approx$1.5 for weak to strong fields), {\it i.e.}, the HMI vs. ~\vsmr\ scaling is more nonlinear. This possibly contributes to the higher
spatial inhomogeneity in Figure \ref{fig:hmi-slope} as compared to Figure \ref{fig:mdi-slope}. Not surprisingly,
the zero-offset, $a$, also increases when strong fluxes are considered.

\begin{figure*}
\includegraphics[width=12cm]{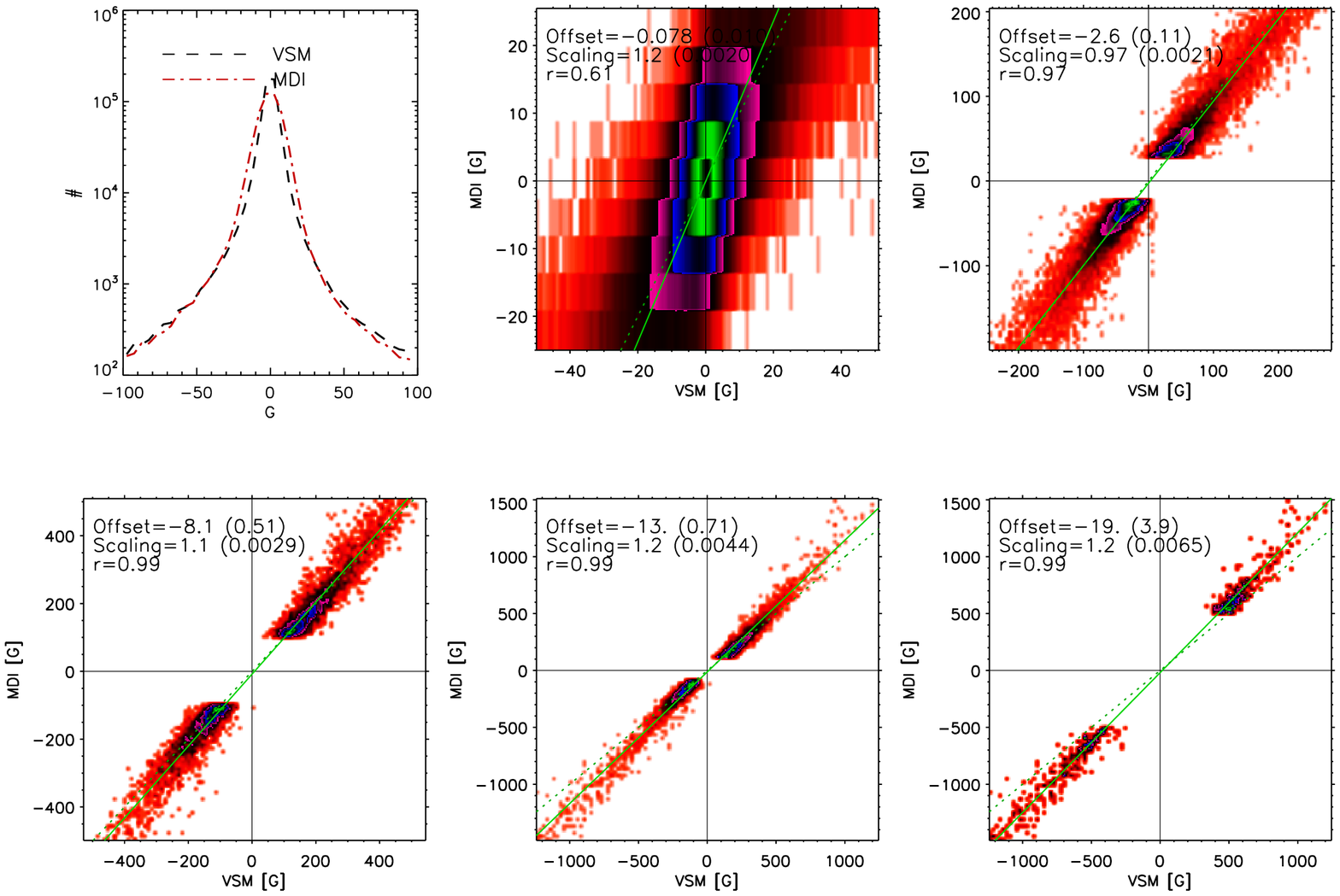}
\caption{Flux dependence of MDI vs. ~\vsmr\ scaling factors. The MDI magnetogram was spatially smoothed with a Gaussian, FWHM=5\arc. Top left:
Histograms for \vsmr\ (black) and MDI (red) data (28 March 2008). Remaining panels show
scatter plots for various flux ranges (color scale for scatter plots
is logarithmic). Shown in solid line is computed fit and in dotted
line scaling factor equal unity and zero offset fit. Computed fit
(scaling factor and offset, standard deviation in fitting in parentheses) and
Pearson correlation coefficient are given at the upper left corner of each
panel. }
\label{fig:mdi-dflux}
\end{figure*}

\begin{figure*}
\includegraphics[width=12cm]{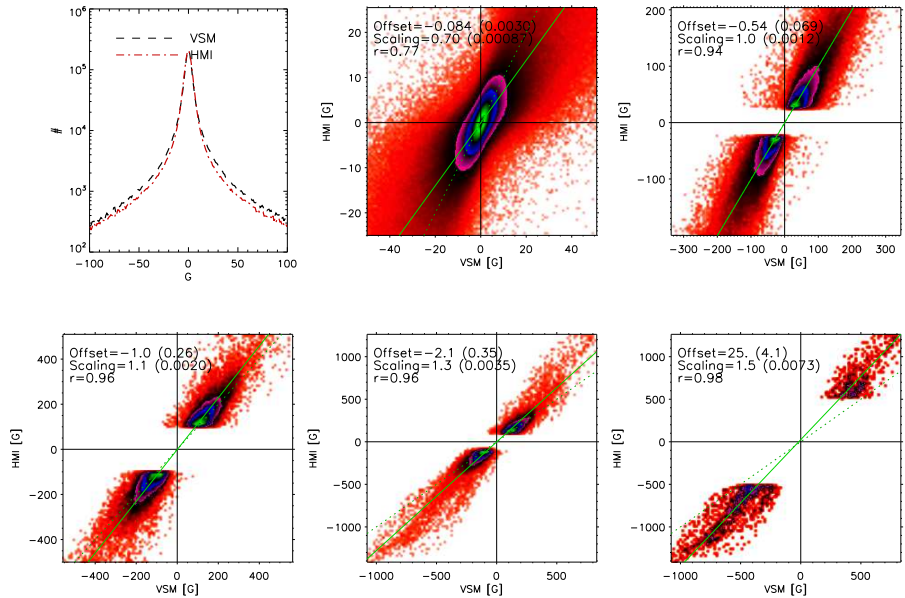}
\caption{Flux dependence of HMI vs. ~\vsms\ scaling factors. The HMI magnetogram was spatially smoothed with a Gaussian, FWHM=4\arc.  Top left:
Histograms for \vsms\ (black) and MDI (red) data (14 November 2011). Remaining panels show
scatter plots for various flux ranges (color scale for scatter plots
is logarithmic). Shown in solid line is computed fit and in dotted
line scaling factor equal unity and zero offset fit. Computed fit
(scaling factor and offset, standard deviation in fitting in parentheses) and
Pearson correlation coefficient are given at the upper left corner of each
panel.}
\label{fig:hmi-dflux}
\end{figure*}

\section{Summary}
Determining scaling factors to inter-calibrate magnetograms from different sources
is challenging and finding universal scaling factors which apply for all flux ranges,
disk positions, spatial resolutions and seeing conditions is unattainable. The comparison presented here, despite the challenges,
demonstrates very good agreement between VSM LOS magnetic flux density measurements
and data from other sources (MWO, SDO/HMI and SOHO/MDI). This is true
both for disk-averaged quantities and pixel-by-pixel comparisons.

Comparison of mean flux density time series shows that the VSM compares
favorably with MWO and MDI data: It has no zero-offset issues found
in both MDI and MWO data. (A zero-offset is present in the first 780 days of the MWO data used in the comparison.) Despite the nonlinearity at high flux densities, the \vsmr\ and
\vsms\ provide a consistent data set of disk-averaged mean flux densities
essential for synoptic studies. 

The comparison showed that the
internal instrument calibration between \vsmr\ and \vsms\ is
not fully linear: The internal calibration coefficient, constructed to reach the
best possible agreement between VSM magnetograms, does not fully reconcile
the mean flux density measurements of \vsmr\ and \vsms\ unless
strong fields are excluded. We do not yet fully understand the reason for this
nonlinearity.  

The noise levels in VSM data are significantly lower than in HMI or MDI,
less than 3 G compared to 10 G and 26 G for HMI and MDI \cite{Liu+others2012}. Even after applying optimal spatial smoothing to HMI and MDI
data, their noise levels remain somewhat higher than those of VSM. We showed
through various analyses how the determination of scaling factors is affected
by spatial smoothing, flux dependence, and center-to-limb variations. Spatial
smoothing is needed to match the apparent flux densities of the different magnetograms.
Smoothing, however, decreases the apparent flux density of the data leading to altered scaling factors. Due to varying seeing conditions, determining optimal smoothing
may not be not trivial when the comparison includes ground-based instruments. While the
correlation of the magnetograms is fairly constant for a wide range of smoothing
amplitudes, the scaling factors can change significantly. Furthermore, for
ground-based instruments the optimal smoothing can change not only from one
observation to another, but, in the worst case, also within an single observation.

The flux dependence of the scaling factors needs to be considered in the
comparisons. In particular, the HMI vs. ~\vsms\ comparison showed that even
in the case of spatially smoothed data, the scaling factors have a non-negligible
dependence on the flux density. There are many possible causes for the nonlinear
scaling. For example, scattered light reduces measured fluxes in sunspots in VSM
observations. Strong blending in the Ni {\sc i} 676.8 nm line in the umbrae of sunspots reduces estimates of flux density in instruments that use this line (MDI and Global Oscillation Network Group, GONG). At low flux densities, the scaling factor between MDI vs. ~\vsmr\ is roughly unity and at high densities the scaling factor increases
to $\approx$1.2. For HMI vs. ~\vsms\ the factors are $\approx$1 for low and$\approx$1.5 for high flux
densities. Note that these values do not address center-to-limb variation and are
computed after spatial smoothing has been applied to the space-based data.

\begin{ack} We thank R. K. Ulrich for providing the MWO mean flux time series. SOLIS data used here are produced cooperatively by NSF/NSO and NASA/LWS. HMI/SDO is a project of NASA. MDI/SOHO is a project of international cooperation between ESA and NASA.

\end{ack}


\begin{thebibliography}{24}
\ifx \bisbn   \undefined \def \bisbn  #1{ISBN #1}\fi
\ifx \binits  \undefined \def \binits#1{#1}\fi
\ifx \bauthor  \undefined \def \bauthor#1{#1}\fi
\ifx \batitle  \undefined \def \batitle#1{#1}\fi
\ifx \bjtitle  \undefined \def \bjtitle#1{\textit{#1}}\fi
\ifx \bvolume  \undefined \def \bvolume#1{\textbf{#1}}\fi
\ifx \byear  \undefined \def \byear#1{#1}\fi
\ifx \bissue  \undefined \def \bissue#1{#1}\fi
\ifx \bfpage  \undefined \def \bfpage#1{#1}\fi
\ifx \blpage  \undefined \def \blpage #1{#1}\fi
\ifx \burl  \undefined \def \burl#1{\textsf{#1}}\fi
\ifx \href  \undefined \def \href#1#2{\textsf{#2}}\fi
\ifx \doiurl  \undefined \def
  \doiurl#1{\href{http://dx.doi.org/#1}{\textsf{#1}}}\fi
\ifx \betal  \undefined \def \betal{\textit{et al.}}\fi
\ifx \binstitute  \undefined \def \binstitute#1{#1}\fi
\ifx \bctitle  \undefined \def \bctitle#1{#1}\fi
\ifx \beditor  \undefined \def \beditor#1{#1}\fi
\ifx \bpublisher  \undefined \def \bpublisher#1{#1}\fi
\ifx \bbtitle  \undefined \def \bbtitle#1{\textit{#1}}\fi
\ifx \bedition  \undefined \def \bedition#1{#1}\fi
\ifx \bseriesno  \undefined \def \bseriesno#1{\textbf{#1}}\fi
\ifx \blocation  \undefined \def \blocation#1{#1}\fi
\ifx \bsertitle  \undefined \def \bsertitle#1{\textit{#1}}\fi
\ifx \bsnm \undefined \def \bsnm#1{#1}\fi
\ifx \bsuffix \undefined \def \bsuffix#1{#1}\fi
\ifx \bparticle \undefined \def \bparticle#1{#1}\fi
\ifx \barticle \undefined \def \barticle#1{}\fi
\ifx \botherref \undefined \def \botherref#1{}\fi
\ifx \url \undefined \def \url#1{\textsf{#1}}\fi
\ifx \bchapter \undefined \def \bchapter#1{}\fi
\ifx \bbook \undefined \def \bbook#1{}\fi
\ifx \bcomment \undefined \def \bcomment#1{#1}\fi
\ifx \oauthor \undefined \def \oauthor#1{#1}\fi
\ifx \citeauthoryear \undefined \def \citeauthoryear#1{#1}\fi
\def \endbibitem {}
\ifx \bconflocation  \undefined \def \bconflocation#1{#1} \fi

\bibitem[\protect\citeauthoryear{{Balasubramaniam} and {Pevtsov}}{2011}]{solis}
\begin{bchapter}
\bauthor{\bsnm{{Balasubramaniam}}, \binits{K.S.}},
\bauthor{\bsnm{{Pevtsov}}, \binits{A.}}:
\byear{2011},
\bctitle{{Ground-based synoptic instrumentation for solar observations}}.
In: Fineschi, S., Fennelly, J. (eds.),\bbtitle{Solar Physics and Space Weather Instrumentation IV.},
\bsertitle{Proc. SPIE}
\bseriesno{8148}.
\bfpage{814809}.
\end{bchapter}
\endbibitem

\bibitem[\protect\citeauthoryear{{Berger} and {Lites}}{2002}]{Berger+Lites2002}
\begin{barticle}
\bauthor{\bsnm{{Berger}}, \binits{T.E.}},
\bauthor{\bsnm{{Lites}}, \binits{B.W.}}:
\byear{2002},
\batitle{{Weak-field magnetogram calibration using Advanced Stokes Polarimeter
  flux-density maps - I. Solar Optical Universal Polarimeter calibration}}.
\bjtitle{\solphys}
\bvolume{208},
\bfpage{181}.
\end{barticle}
\endbibitem

\bibitem[\protect\citeauthoryear{{Berger} and {Lites}}{2003}]{Berger+Lites2003}
\begin{barticle}
\bauthor{\bsnm{{Berger}}, \binits{T.E.}},
\bauthor{\bsnm{{Lites}}, \binits{B.W.}}:
\byear{2003},
\batitle{{Weak-field magnetogram calibration using Advanced Stokes Polarimeter
  flux density maps - II. SOHO/MDI full-disk mode calibration}}.
\bjtitle{\solphys}
\bvolume{213},
\bfpage{213}.
\end{barticle}
\endbibitem

\bibitem[\protect\citeauthoryear{{Cauzzi}
  \textit{et~al.}}{1993}]{Cauzzi+others1993}
\begin{barticle}
\bauthor{\bsnm{{Cauzzi}}, \binits{G.}},
\bauthor{\bsnm{{Smaldone}}, \binits{L.A.}},
\bauthor{\bsnm{{Balasubramaniam}}, \binits{K.S.}},
\bauthor{\bsnm{{Keil}}, \binits{S.L.}}:
\byear{1993},
\batitle{{On the calibration of line-of-sight magnetograms}}.
\bjtitle{\solphys}
\bvolume{146},
\bfpage{207}.
\end{barticle}
\endbibitem

\bibitem[\protect\citeauthoryear{{Demidov}
  \textit{et~al.}}{2008}]{Demidov+others2008}
\begin{barticle}
\bauthor{\bsnm{{Demidov}}, \binits{M.L.}},
\bauthor{\bsnm{{Golubeva}}, \binits{E.M.}},
\bauthor{\bsnm{{Balthasar}}, \binits{H.}},
\bauthor{\bsnm{{Staude}}, \binits{J.}},
\bauthor{\bsnm{{Grigoryev}}, \binits{V.M.}}:
\byear{2008},
\batitle{{Comparison of solar magnetic fields measured at different
  observatories: peculiar strength ratio distributions across the disk}}.
\bjtitle{\solphys}
\bvolume{250},
\bfpage{279}.
\end{barticle}
\endbibitem

\bibitem[\protect\citeauthoryear{{Freeland} and {Handy}}{1998}]{SSW}
\begin{barticle}
\bauthor{\bsnm{{Freeland}}, \binits{S.L.}},
\bauthor{\bsnm{{Handy}}, \binits{B.N.}}:
\byear{1998},
\batitle{{Data analysis with the SolarSoft System}}.
\bjtitle{\solphys}
\bvolume{182},
\bfpage{497}.
\end{barticle}
\endbibitem

\bibitem[\protect\citeauthoryear{{Harvey}
  \textit{et~al.}}{2007}]{Harvey+others2007}
\begin{barticle}
\bauthor{\bsnm{{Harvey}}, \binits{J.W.}},
\bauthor{\bsnm{{Branston}}, \binits{D.}},
\bauthor{\bsnm{{Henney}}, \binits{C.J.}},
\bauthor{\bsnm{{Keller}}, \binits{C.U.}},
\bauthor{\bsnm{{SOLIS and GONG Teams}}}:
\byear{2007},
\batitle{{Seething horizontal magnetic fields in the quiet solar photosphere}}.
\bjtitle{\apjl}
\bvolume{659},
\bfpage{L177}.
\end{barticle}
\endbibitem

\bibitem[\protect\citeauthoryear{{Isobe} \textit{et~al.}}{1990}]{sixlin}
\begin{barticle}
\bauthor{\bsnm{{Isobe}}, \binits{T.}},
\bauthor{\bsnm{{Feigelson}}, \binits{E.D.}},
\bauthor{\bsnm{{Akritas}}, \binits{M.G.}},
\bauthor{\bsnm{{Babu}}, \binits{G.J.}}:
\byear{1990},
\batitle{{Linear regression in astronomy.}}
\bjtitle{\apj}
\bvolume{364},
\bfpage{104}.
\end{barticle}
\endbibitem

\bibitem[\protect\citeauthoryear{{Jones}}{1996}]{Jones1996}
\begin{bchapter}
\bauthor{\bsnm{{Jones}}, \binits{H.P.}}:
\byear{1996},
\bctitle{{Online analysis and compression of spectra-spectroheliograms}}.
In: \beditor{\bsnm{{Rust}},\binits{D.M.}} (ed.).
\bbtitle{Missions to the Sun},
\bsertitle{Proc. SPIE}
\bseriesno{2804},
\bfpage{110}.
\end{bchapter}
\endbibitem

\bibitem[\protect\citeauthoryear{{Jones} and {Ceja}}{2001}]{Jones+Ceja2001}
\begin{bchapter}
\bauthor{\bsnm{{Jones}}, \binits{H.P.}},
\bauthor{\bsnm{{Ceja}}, \binits{J.A.}}:
\byear{2001},
\bctitle{{Preliminary comparison of magnetograms from KPVT/SPM, SOHO/MDI and
  GONG$^{+}$}}.
In: \beditor{\bsnm{{Sigwarth}}, \binits{M.}} (ed.)
\bbtitle{Advanced Solar Polarimetry -- Theory, Observation, and
  Instrumentation},
\bsertitle{ASP. Conf. Ser.}
\bseriesno{236},
\bfpage{87}.
\end{bchapter}
\endbibitem

\bibitem[\protect\citeauthoryear{{Jones}
  \textit{et~al.}}{1992}]{Jones+others1992}
\begin{barticle}
\bauthor{\bsnm{{Jones}}, \binits{H.P.}},
\bauthor{\bsnm{{Duvall}}, \binits{T.L.} \bsuffix{Jr.}},
\bauthor{\bsnm{{Harvey}}, \binits{J.W.}},
\bauthor{\bsnm{{Mahaffey}}, \binits{C.T.}},
\bauthor{\bsnm{{Schwitters}}, \binits{J.D.}},
\bauthor{\bsnm{{Simmons}}, \binits{J.E.}}:
\byear{1992},
\batitle{{The NASA/NSO spectromagnetograph}}.
\bjtitle{\solphys}
\bvolume{139},
\bfpage{211}.
\end{barticle}
\endbibitem

\bibitem[\protect\citeauthoryear{{Jones}
  \textit{et~al.}}{2004}]{Jones+others2004}
\begin{barticle}
\bauthor{\bsnm{{Jones}}, \binits{H.P.}},
\bauthor{\bsnm{{Harvey}}, \binits{J.W.}},
\bauthor{\bsnm{{Henney}}, \binits{C.J.}},
\bauthor{\bsnm{{Keller}}, \binits{C.U.}},
\bauthor{\bsnm{{Malanushenko}}, \binits{O.M.}}:
\byear{2004},
\batitle{{Measurement scale of the SOLIS vector spectromagnetograph}}.
\bjtitle{Bull. Am. Astron. Soc.}
\bvolume{36},
\bfpage{709}.
\end{barticle}
\endbibitem

\bibitem[\protect\citeauthoryear{{Keller}, {Harvey}, and
  {Giampapa}}{2003}]{Keller+others2003}
\begin{bchapter}
\bauthor{\bsnm{{Keller}}, \binits{C.U.}},
\bauthor{\bsnm{{Harvey}}, \binits{J.W.}},
\bauthor{\bsnm{{Giampapa}}, \binits{M.S.}}:
\byear{2003},
\bctitle{{SOLIS: An innovative suite of synoptic instruments}}.
In: \beditor{\bsnm{{Keil, S.~L., Avakyan, S.~V.}}} (eds.)
\bbtitle{Innovative Telescopes and Instrumentation for Solar Astrophysics},
\bsertitle{Proc. SPIE}
\bseriesno{4853},
\bfpage{194}.
\end{bchapter}
\endbibitem

\bibitem[\protect\citeauthoryear{{Leka} and {Barnes}}{2012}]{Leka+Barnes2012}
\begin{barticle}
\bauthor{\bsnm{{Leka}}, \binits{K.D.}},
\bauthor{\bsnm{{Barnes}}, \binits{G.}}:
\byear{2012},
\batitle{{Modeling and interpreting the effects of spatial resolution on solar
  magnetic field maps}}.
\bjtitle{\solphys}
\bvolume{277},
\bfpage{89}.
\end{barticle}
\endbibitem

\bibitem[\protect\citeauthoryear{{Liu} \textit{et~al.}}{2012}]{Liu+others2012}
\begin{barticle}
\bauthor{\bsnm{{Liu}}, \binits{Y.}},
\bauthor{\bsnm{{Hoeksema}}, \binits{J.T.}},
\bauthor{\bsnm{{Scherrer}}, \binits{P.H.}},
\bauthor{\bsnm{{Schou}}, \binits{J.}},
\bauthor{\bsnm{{Couvidat}}, \binits{S.}},
\bauthor{\bsnm{{Bush}}, \binits{R.I.}},
\bauthor{\bsnm{{Duvall}}, \binits{T.L.}},
\bauthor{\bsnm{{Hayashi}}, \binits{K.}},
\bauthor{\bsnm{{Sun}}, \binits{X.}},
\bauthor{\bsnm{{Zhao}}, \binits{X.}}:
\byear{2012},
\batitle{{Comparison of line-of-sight magnetograms taken by the Solar Dynamics
  Observatory/Helioseismic and Magnetic Imager and Solar and Heliospheric
  Observatory/Michelson Doppler Imager}}.
\bjtitle{\solphys}
\bvolume{279},
\bfpage{295}.
\end{barticle}
\endbibitem

\bibitem[\protect\citeauthoryear{{Livingston} \textit{et~al.}}{1976a}]{kpvt1}
\begin{barticle}
\bauthor{\bsnm{{Livingston}}, \binits{W.C.}},
\bauthor{\bsnm{{Harvey}}, \binits{J.}},
\bauthor{\bsnm{{Pierce}}, \binits{A.K.}},
\bauthor{\bsnm{{Schrage}}, \binits{D.}}
\bauthor{\bsnm{{Gillespie}}, \binits{B.}},
\bauthor{\bsnm{{Simmons}}, \binits{J.}},
\bauthor{\bsnm{{Slaughter}}, \binits{C.}}:
\byear{1976}a,
\batitle{{Kitt Peak 60-cm vacuum telescope}}.
\bjtitle{Appl. Opt.}
\bvolume{15},
\bfpage{33}.
\end{barticle}
\endbibitem

\bibitem[\protect\citeauthoryear{{Livingston} \textit{et~al.}}{1976b}]{kpvt2}
\begin{barticle}
\bauthor{\bsnm{{Livingston}}, \binits{W.C.}},
\bauthor{\bsnm{{Harvey}}, \binits{J.}},
\bauthor{\bsnm{{Slaughter}}, \binits{C.}},
\bauthor{\bsnm{{Trumbo}}, \binits{D.}}:
\byear{1976}b,
\batitle{{Solar magnetograph employing integrated diode arrays}}.
\bjtitle{App. Opt.}
\bvolume{15},
\bfpage{40}.
\end{barticle}
\endbibitem

\bibitem[\protect\citeauthoryear{{Rees} and {Semel}}{1979}]{Rees+Semel1979}
\begin{barticle}
\bauthor{\bsnm{{Rees}}, \binits{D.E.}},
\bauthor{\bsnm{{Semel}}, \binits{M.D.}}:
\byear{1979},
\batitle{{Line formation in an unresolved magnetic element - A test of the
  centre of gravity method}}.
\bjtitle{\aap}
\bvolume{74},
\bfpage{1}.
\end{barticle}
\endbibitem

\bibitem[\protect\citeauthoryear{{Scherrer} \textit{et~al.}}{1995}]{mdi}
\begin{barticle}
\bauthor{\bsnm{{Scherrer}}, \binits{P.H.}},
\bauthor{\bsnm{{Bogart}}, \binits{R.S.}},
\bauthor{\bsnm{{Bush}}, \binits{R.I.}},
\bauthor{\bsnm{{Hoeksema}}, \binits{J.T.}},
\bauthor{\bsnm{{Kosovichev}}, \binits{A.G.}},
\bauthor{\bsnm{{Schou}}, \binits{J.}},
\betal
\byear{1995},
\batitle{{The Solar Oscillations Investigation - Michelson Doppler Imager}}.
\bjtitle{\solphys}
\bvolume{162},
\bfpage{129}.
\end{barticle}
\endbibitem

\bibitem[\protect\citeauthoryear{{Scherrer} \textit{et~al.}}{2012}]{HMI}
\begin{barticle}
\bauthor{\bsnm{{Scherrer}}, \binits{P.H.}},
\bauthor{\bsnm{{Schou}}, \binits{J.}},
\bauthor{\bsnm{{Bush}}, \binits{R.I.}},
\bauthor{\bsnm{{Kosovichev}}, \binits{A.G.}},
\bauthor{\bsnm{{Bogart}}, \binits{R.S.}},
\bauthor{\bsnm{{Hoeksema}}, \binits{J.T.}},
\betal\byear{2012},
\batitle{{The Helioseismic and Magnetic Imager (HMI) Investigation for the S
  olar Dynamics Observatory (SDO)}}.
\bjtitle{\solphys}
\bvolume{275},
\bfpage{207}.
\end{barticle}
\endbibitem

\bibitem[\protect\citeauthoryear{{Thornton} and
  {Jones}}{2002}]{Thornton+Jones2002}
\begin{barticle}
\bauthor{\bsnm{{Thornton}}, \binits{C.E.}},
\bauthor{\bsnm{{Jones}}, \binits{H.P.}}:
\byear{2002},
\batitle{{Comparison of three solar magnetographs}}.
\bjtitle{Bull. Am. Astron. Soc.}
\bvolume{34},
\bfpage{1243}.
\end{barticle}
\endbibitem

\bibitem[\protect\citeauthoryear{{Tran}
  \textit{et~al.}}{2005}]{Tran+others2005}
\begin{barticle}
\bauthor{\bsnm{{Tran}}, \binits{T.}},
\bauthor{\bsnm{{Bertello}}, \binits{L.}},
\bauthor{\bsnm{{Ulrich}}, \binits{R.K.}},
\bauthor{\bsnm{{Evans}}, \binits{S.}}:
\byear{2005},
\batitle{{Magnetic fields from SOHO MDI converted to the Mount Wilson 150 Foot
  Solar Tower scale}}.
\bjtitle{\apjs}
\bvolume{156},
\bfpage{295}.
\end{barticle}
\endbibitem

\bibitem[\protect\citeauthoryear{{Ulrich}
  \textit{et~al.}}{2002}]{Ulrich+others2002}
\begin{barticle}
\bauthor{\bsnm{{Ulrich}}, \binits{R.K.}},
\bauthor{\bsnm{{Evans}}, \binits{S.}},
\bauthor{\bsnm{{Boyden}}, \binits{J.E.}},
\bauthor{\bsnm{{Webster}}, \binits{L.}}:
\byear{2002},
\batitle{{Mount Wilson synoptic magnetic fields: Improved instrumentation,
  calibration, and analysis applied to the 2000 July 14 flare and to the
  evolution of the dipole field}}.
\bjtitle{\apjs}
\bvolume{139},
\bfpage{259}.
\end{barticle}
\endbibitem

\bibitem[\protect\citeauthoryear{{Wenzler}
  \textit{et~al.}}{2004}]{Wenzler+others2004}
\begin{barticle}
\bauthor{\bsnm{{Wenzler}}, \binits{T.}},
\bauthor{\bsnm{{Solanki}}, \binits{S.K.}},
\bauthor{\bsnm{{Krivova}}, \binits{N.A.}},
\bauthor{\bsnm{{Fluri}}, \binits{D.M.}}:
\byear{2004},
\batitle{{Comparison between KPVT/SPM and SoHO/MDI magnetograms with an
  application to solar irradiance reconstructions}}.
\bjtitle{\aap}
\bvolume{427},
\bfpage{1031}.
\end{barticle}
\endbibitem

\end{thebibliography}

\end{article} 

\end{document}